\documentclass[12pt]{article}

\usepackage{graphicx}
\usepackage{epsfig}

\begin{document}
\begin{center}
{\Large\bf Production of $\varphi$-mesons on nuclear targets\\ in the Quark-Gluon String model} \\ 
\vspace{0.6cm}

G.H.~Arakelyan$^1$, C.~Merino$^2$, Yu.M.~Shabelski$^{3}$ \\

\vspace{.5cm}
$^1$A.Alikhanyan National Scientific Laboratory \\
(Yerevan Physics Institute)\\
Yerevan, 0036, Armenia\\
e-mail: argev@mail.yerphi.am\\
\vspace{0.1cm}

$^2$Departamento de F\'\i sica de Part\'\i culas, Facultade de F\'\i sica\\
and Instituto Galego de F\'\i sica de Altas Enerx\'\i as (IGFAE)\\
Universidade de Santiago de Compostela\\
15782 Santiago de Compostela\\
Galiza-Spain\\
e-mail: merino@fpaxp1.usc.es \\
\vspace{0.1cm}

$^3$Petersburg Nuclear Physics Institute\\
NCR Kurchatov Institute\\
Gatchina, St.Petersburg 188350, Russia\\
e-mail: shabelsk@thd.pnpi.spb.ru
\vskip 0.9 truecm

\vspace{1.2cm}

{\bf Abstract}
\end{center}

We consider the experimental data on $\varphi$-meson production on nuclear targets,
and we find that they present unusually small shadow corrections
for the inclusive density in the midrapidity region.
We also give a quantitatively consistent description of
both the initial energy dependence and the A-dependence of the produced $\varphi$-mesons,
obtained in the frame of the Quark-Gluon String Model.

\vskip 1.5cm


\newpage

%
%{\Large \bf
\vskip -1.5cm
\section{Introduction}

The experimental data on the production of $\varphi$-mesons, a rarely
produced system formed of $s\overline{s}$ quarks,
%%%with non-zero masses,
in proton-nucleus and nucleus-nucleus collisions show at very high energies
unusually small shadow correction effects in the midrapididty region.

In this paper we discuss a possible reason for this effect that takes into account
the fact that though $s$ and $\overline{s}$ quarks have non-zero masses,
at the same time their masses are not large enough to make standard perturbative
Quantum Chromodynamics applicable, what could allow the treatment of the $\varphi$-meson
as an intermediate case between soft and hard physics. We present
the corresponding theoretical results, obtained by using the formalism of the
Quark-Gluon String Model (QGSM)~\cite{KTM,K20}.

This model has been successfully used for the description of
multiparticle production processes in hadron-nucleon~\cite{KTM,K20},
hadron-nucleus~\cite{KTMS,Sh1} and nucleus-nucleus~\cite{Sha,AMPSpl} 
collisions. The QGSM description of the production of secondary pseudoscalar
mesons $\pi$ and $K$, and of baryons $p$, $\overline{p}$, $\Lambda$, and $\overline{\Lambda}$,
was obtained many years ago in~\cite{KaPi,Sh,AMPS,AMPSk}. Vector meson production was 
considered in~\cite{aryer,yer,APSh,amsphi}. The yields of hyperons, including the 
multistrange ones, has been described in~\cite{ACKS,Sigma}.

In the case of collisions with a nuclear target,
the saturation of the inclusive density of secondaries~\cite{ACKS}
found at very high energies was also
successfully described by QGSM~\cite{ACKS,MPSppb,MPSd,MPSpa}.

Already in the paper~\cite{amsphi} we had for the first time applied the QGSM formalism
to the description of the spectra  of vector $\varphi$-mesons production
in $\pi p$ and $pp$ collisions, and of the ratios 
of yields $\varphi$/$\pi^-$ and $\varphi$/$K^-$ in $pp$ collisions for a large scope of the 
initial energy, going up to the RHIC and LHC ranges.

In this paper we extend our analysis to the production of $\varphi$-mesons
in high-energy collisions on a nuclear target, and we provide a consistent
description of the small shadow correction effects measured for this case
in the midrapididty region.
%%%without introducing any additional parameter
%%%with respect to the description of $\varphi$-mesons production given for pp collisions.

\section{Meson inclusive spectra in the QGSM}

In order to produce quantitative results for the integrated over $p_T$
inclusive spectra of secondary hadrons,
a model for multiparticle production is needed. It is for that purpose that we have
used the QGSM~\cite{KTM,K20} in the numerical calculations presented below.

The QGSM~\cite{KTM,K20}, based on the Dual
Topological Unitarization,
%%%(DTU),
Regge phenomenology, and nonperturbative
notions of QCD, has been used for already more than thirty years
to succesfully predict and describe many features of the hadronic processes
in a wide energy range. In particular, the QGSM allows one to make
quantitative predictions on the inclusive densities of different secondaries
produced at high energy collisions both in the central and beam fragmentation regions.

In the QGSM, the high energy hadron-nucleon, hadron-nucleus, and nucleus-nucleus 
interactions are treated as proceeding via the exchange of one or several Pomerons,
and all elastic and inelastic processes result from cutting through or between
Pomerons~\cite{AGK}. Each Pomeron corresponds to a cylinder diagram (see
Fig.~1a), in which the cylinder boundaries are drawn by the dash-dotted
vertical lines. The surface of the cylinder is schematically
depicted by dashed lines, while the solid lines at the top and bottom
of the cylinder represent, respectively, the beam and the target quarks,
which interaction is mediated by the Pomeron exchange.

%and nucleus-nucleus

%
% Thy cylindrical surface in Fig1a is schematically shown
% by dashed lines
% and
% the boundaries of cylinder by dash-dotted curves.
% The beam and target quarks which interacts via exchange
% by this cylinder, are shown by solid curves.
%
% that, when cut,
The cut through the cylinder produces two showers of secondaries, i.e.
%
% as it is shown in Fig.~1b.
% The newly produced
quark-antiquark pairs shown in Fig~1b by solid lines.
The inclusive spectrum of secondaries
is then determined
by the convolution of diquark, valence quark, and sea quark distributions
in the incident particles, $u(x,n)$, with the fragmentation functions
of quarks and diquarks into the secondary hadrons, $G(z)$.
Both functions $u(x,n)$ and $G(z)$ are
determined by the appropriate Reggeon diagrams~\cite{Kai}.
Note that the quark-antiquark distributions $u(x,n)$ differ
from the standard PDF's extracted from fits to experimental data
because the $u(x,n)$'s are theoretically taken to be valid 
at the rather low $Q^2$ which are relevant for soft processes,
while the PDF distributions are obtained by fixing the
behavior at large $Q^2$.
%
% these functions describe the distribution
% at rather low $Q^2$ which are needed for the description
% of soft processes whereas the PDF distributions are related
% to behavior at large $Q^2$.
The diquark and quark distribution functions depend
on the number $n$ of cut
Pomerons in the considered diagram.
For the following calculations we have used the
prescription given in reference~\cite{KTMS}.

\begin{figure}[htb]
%%%\centering
%%%\vskip -2.cm
%%%\hskip 2.cm
\vskip -5.5cm
\hskip 4.cm
\includegraphics[width=.6\hsize]{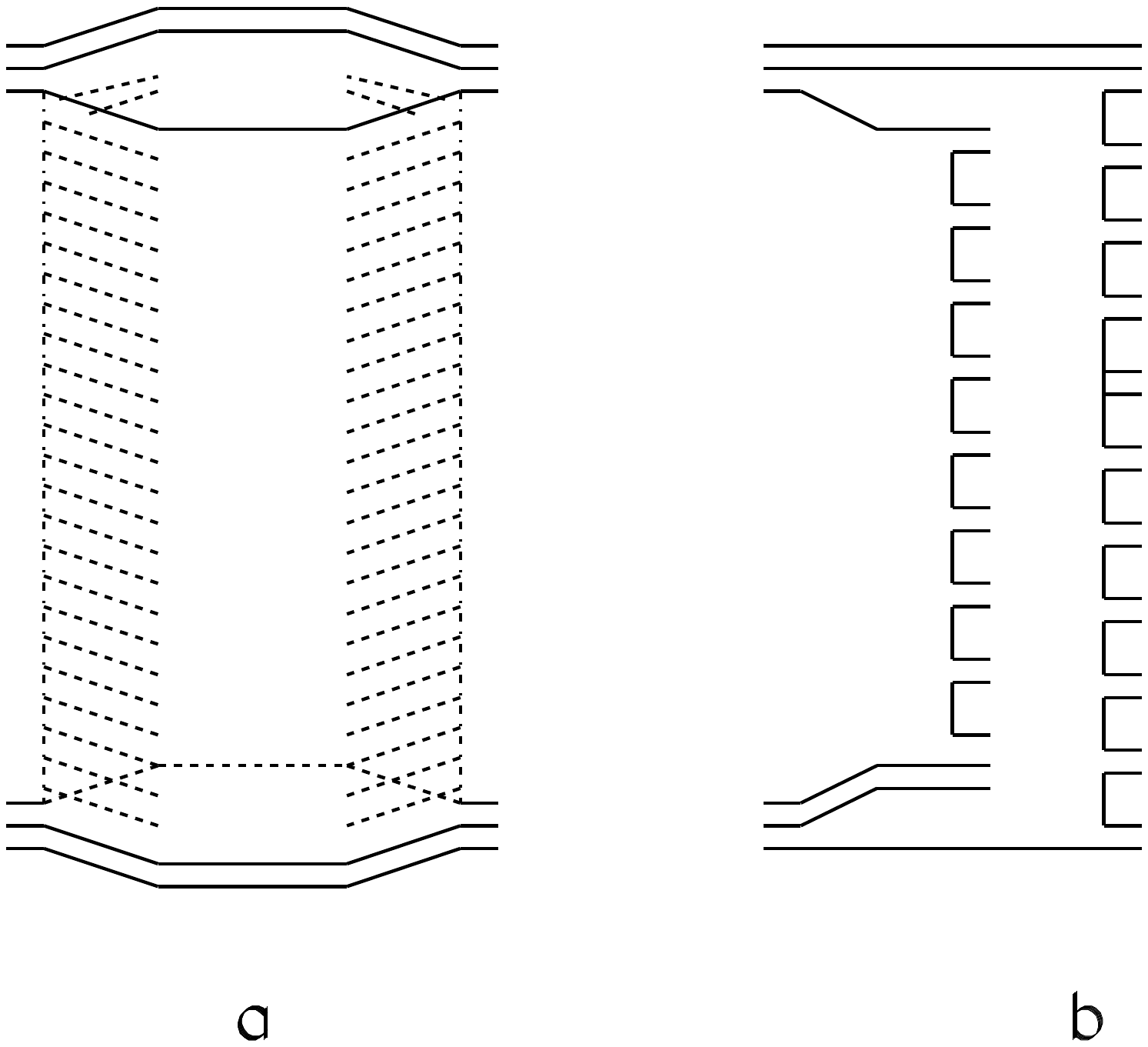}
%%%\includegraphics[width=.55\hsize]{fi00.eps}
%%%\vskip -2.5cm
\vskip -6.cm
\hskip 3.cm
\includegraphics[width=.6\hsize]{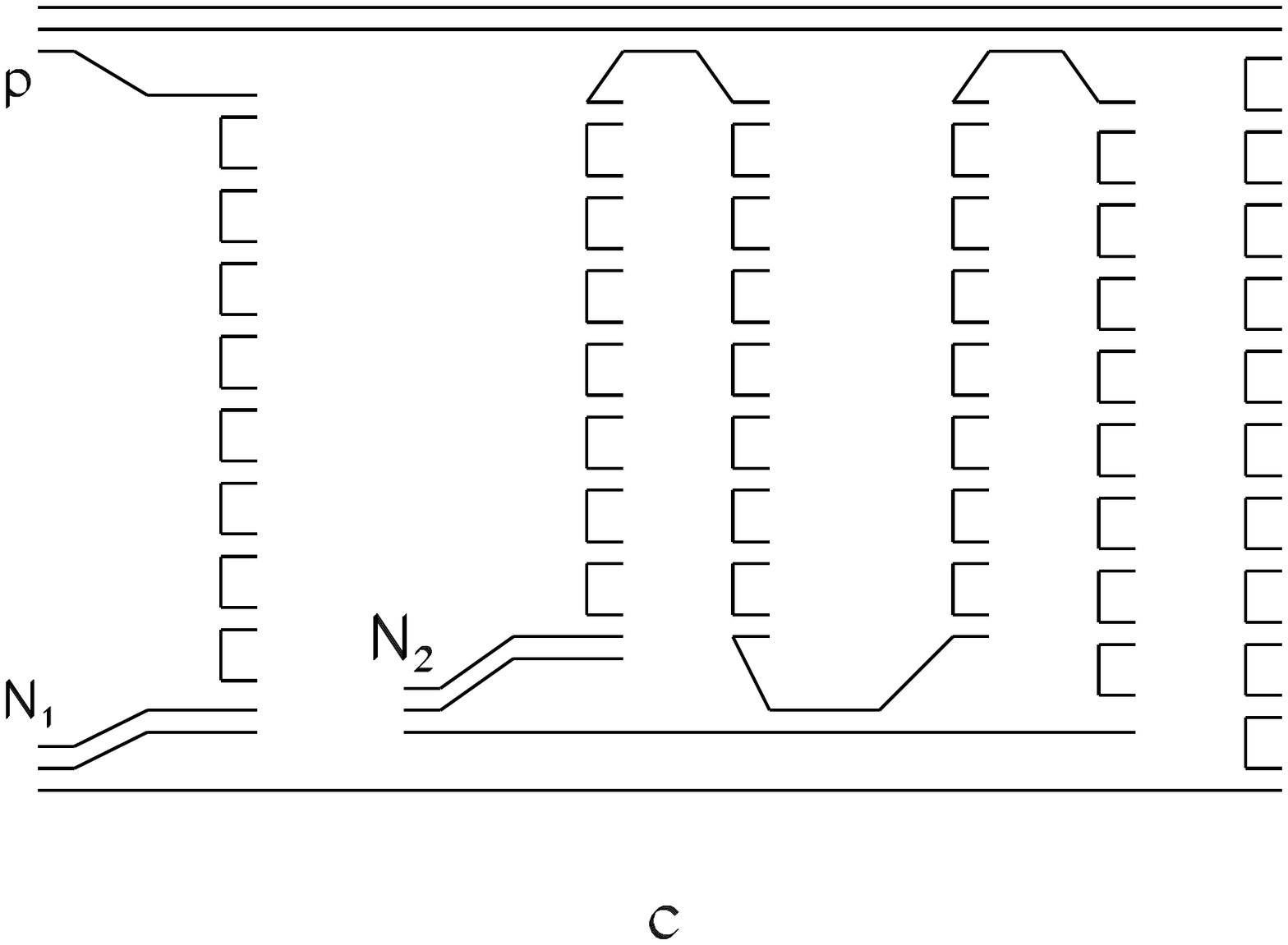}
\vskip -1.cm
\caption{\footnotesize
(a) Cylindrical diagram representing the Pomeron exchange
within the Dual Topological Unitarization (DTU) classification
(quarks are shown by solid lines);
(b) Cut of the cylindrical diagram corresponding to the single-Pomeron
exchange contribution in inelastic $pp$ scattering;
(c) Diagram corresponding to the inelastic interaction
of an incident proton with two target nucleons $N_1$ and $N_2$
in a $pA$ collision.}
\end{figure}

%In the QGSM the inclusive spectrum of a secondary hadron $h$ is
%determined~\cite{KTM,K20} by the convolution of the diquark, valence quark, and %sea 
%quark distributions, $u(x,n)$, in the incident particles, with the 
%fragmentation functions, $G^h(z)$, of quarks and diquarks into the secondary %hadron $h$.
%Both the distribution and the fragmentation functions are constructed using the %Reggeon counting 
%rules~\cite{Kai}.

For a nucleon target, the inclusive rapidity, $y$, or Feynman-$x$, $x_F$,
spectrum of a secondary hadron $h$ has the form~\cite{KTM}:
\begin{equation}
\frac{dn}{dy}\ = \
\frac{x_E}{\sigma_{inel}}\cdot \frac{d\sigma}{dx_F}\ = 
\sum_{n=1}^\infty w_n\cdot\phi_n^h (x) \ ,
\end{equation}
where the functions $\phi_{n}^{h}(x)$ determine the contribution of diagrams
with $n$ cut Pomerons and $w_n$ is the relative weight of this diagram.  
Here, for the $\varphi$-meson production in the midrapidity region,
we neglect the contribution of diffraction and dissociation processes.

In the case of $pp$ collisions:
\begin{eqnarray}
\phi_n^{h}(x) &=& f_{qq}^{h}(x_{+},n) \cdot f_{q}^{h}(x_{-},n) +
f_{q}^{h}(x_{+},n) \cdot f_{qq}^{h}(x_{-},n) \nonumber\\
&+& 2(n-1)\cdot f_{s}^{h}(x_{+},n) \cdot f_{s}^{h}(x_{-},n)\ \  , \\
x_{\pm} &=& \frac{1}{2}[\sqrt{4m_{T}^{2}/s+x^{2}}\pm{x}]\ \ , 
\end{eqnarray}
where $f_{qq}$, $f_{q}$, and $f_{s}$ correspond to the contributions
of diquarks, valence quarks, and sea quarks, respectively.

These contributions are determined by the convolution of the diquark and
quark distributions with the fragmentation functions, e.g.,

\begin{equation}
f_{q}^{h}(x_{+},n) = \int_{x_{+}}^{1}
u_{q}(x_{1},n)G_{q}^{h}(x_{+}/x_{1}) dx_{1}\ \ .
\end{equation}
%The diquark and quark distributions, as well as the fragmentation
%functions, are determined by Regge intercepts \cite{Kai}.

In the calculation of the inclusive spectra of secondaries produced
in $pA$ collisions we should consider the possibility of one
or several Pomeron cuts in each of the $\nu$ blobs of the proton-nucleon
inelastic interactions.
For example, in Fig.~1c it is shown one of the diagrams contributing
to the inelastic interaction of a beam proton with two nucleons from the target.
In the blob of the proton-nucleon$_1$ interaction one Pomeron is cut,
and in the blob of the proton-nucleon$_2$ interaction two Pomerons
are cut. It is essential to take into account all the diagrams with every
possible Pomeron configuration and its permutations. The diquark
and quark distributions and the fragmentation functions here
are the same as in the case of the interaction
with one nucleon.

The process shown in Fig.~1c satisfies~\cite{Sh3,BT,Weis,Jar}
the condition that the absorptive parts of the hadron-nucleus
amplitude are determined by the combination of the absorptive parts
of the hadron-nucleon amplitudes.

In the case of a nucleus-nucleus collision, in the fragmentation region of
the projectile we use the approach~\cite{Sha,Shab,JDDS}, where the beam of
independent nucleons of the projectile interacts with the target nucleus,
what corresponds to the rigid target approximation~\cite{Alk} of
Glauber Theory. In the target fragmentation region, on the contrary, the
beam of independent target nucleons interacts with the projectile nucleus,
these two contributions coinciding in the central region. The corrections for energy
conservation play here a very important role if the initial energy is not
very high. This approach was used in~\cite{JDDS} for the successful
description of $\pi^{\pm}$, $K^{\pm}$, $p$, and $\overline{p}$ produced in PbPb
collisions at 158 GeV/c per nucleon.

%The details of the model are presented in~\cite{KTM,K20,KaPi,Sh,ACKS}, and 
We use in this paper the values of the Pomeron parameters in ref.~\cite{Sh},
and the fragmentation functions of quarks and diquarks into $\varphi$-meson
are presented in ref.~\cite{amsphi}.

%The averaged number of exchanged Pomerons in $pp$ collisions
%$\langle n \rangle_{pp}$ slowly increases with the energy.
%In particular, in the case of $n > 1$, i.e. in the 
%case of multipomeron exchange, the distributions of valence quarks and 
%diquarks are softened due to the appearance of a sea quark contribution 
%\cite{KTMS}. 

%For the $\varphi$-meson production we use the following quark fragmentation
%functions~\cite{aryer}:
%\begin{eqnarray}
%G_{u}^{\varphi} &=& G_{d}^{\varphi} =
%a_{\varphi}\cdot(1-z)^{\lambda - \alpha_R - 2 \alpha_{\varphi}+2}, \;\; \\ 
% \\ \nonumber,
%G_{s}^{\varphi} &=& a_{\varphi}\cdot(1-z)^{\lambda - \alpha_{\varphi}}.
%\end{eqnarray}
%On the other hand, the diquark fragmentation functions into $\varphi$-mesons %have the form:
%\begin{equation}
%G_{uu}^{\varphi} = G_{ud}^{\varphi} = 
%a_{\varphi}\cdot(1-z)^{\lambda+\alpha_R -2(\alpha_R +\alpha_{\varphi})} \ \ ,
%\end{equation}
%where the parameter $\lambda$ takes the value $\lambda$=0.5~\cite{KTM,KaPi,Sh}, %and the
%parameters $\alpha_R$=0.5 and $\alpha_{\varphi}$=0. are the intercepts of the %$\rho$
%and $\varphi$ Regge trajectories, respectively.
%The value of the parameter $a_{\varphi}$ is determined by comparison to %experimental 
%data on $\phi$ production from different hadron beams. In our calcualtions we
%use the value $a_{\varphi}=0.11$.

\section{The $\varphi$-meson production in pA collisions}

We will start from the case of $\varphi$-meson production in pA collisions.
The experimental data obtained by the HERAb Collaboration
at $\sqrt{s}=41.6$ GeV~\cite{HERAb} are presented in Table~1 and Fig.~2.
\vskip 0.5cm

\begin{center}
\vskip -10pt
\begin{tabular}{|c|c|c|} \hline
Reactions & Experimental data & QGSM \\
%\cline{3-5}
% particles &CMS Collaboration & QGSM  & QGSM &  \\
   &  $d\sigma_{pA}/dy$, $|y|\leq 0.5$  &  \\
\hline
%%%\hline

p + C & 1.74 $\pm$ 0.15  & 1.5 \\ \hline

p + Ti & 6.85 $\pm$ 0.7 & 7.1 \\ \hline

p + W & 23.5 $\pm$ 2.1 & 19.1  \\ \hline
%%%\hline

%%%\hline
\end{tabular}
\end{center}
Table~1: {\footnotesize The experimental data for $\varphi$-mesons
production in pA collisions at $\sqrt{s}=41.6$ GeV
by the HERAb Collaboration~\cite{HERAb}, together with the corresponding
QGSM results.}

Due to the AGK cancel rules~\cite{AGK}, the inclusive cross section $d\sigma/dy$
in midrapidity region should present a linear A-dependence:
\begin{equation} 
d\sigma_{pA}/dy (y \simeq 0) = A \cdot d\sigma_{pp}/dy (y \simeq 0)\;. 
\end{equation}

Some numerically small corrections to this linear behaviour are connected to the
energy conservation rule when the initial energy is not high enough~\cite{Sh3}.
One can see in Fig.~2 that these corrections are really small at
the energy $\sqrt{s}=41.6$~GeV (the difference between the solid and the dashed
curves is small for a very large range of $A$).

%\newpage
\begin{figure}[htb]
\label{a}
%%%\centering
\vskip -9.cm
\hskip 3.cm
\includegraphics[width=0.9\hsize]{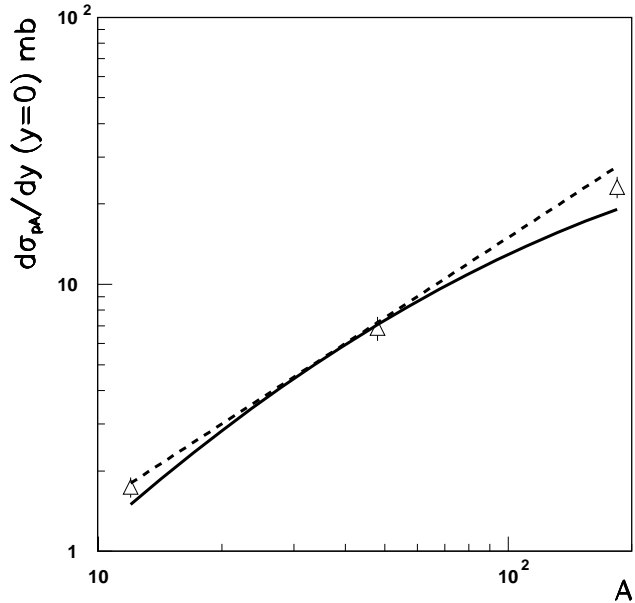}
\vskip -.8cm
\caption{\footnotesize
The experimental data on the A-dependence of $d\sigma_{pA}/dy (y \simeq 0)$ of produced
$\varphi$-mesons in pA collisions at $\sqrt{s}$ = 41.6 Gev~\cite{HERAb},
together with the corresponding QGSM results (solid curve),
and with the linear dependence $d\sigma/dy \propto A^1$ (dashed straight line).}
\end{figure}

Also in ref.~\cite{HERAb}, the rapidity distribution of inclusive cross section
for $\varphi$-mesons production in pA collisions are presented for rather small
rapidity ranges. The existing experimental points, together with the results of
the QGSM calculations, are shown in Fig.~3. The comparison between theory and experiment
seems to be good.

\begin{figure}[htb]
\label{phi400}
%%%%\centering
\vskip -9.cm
\hskip 3.cm
\includegraphics[width=.9\hsize]{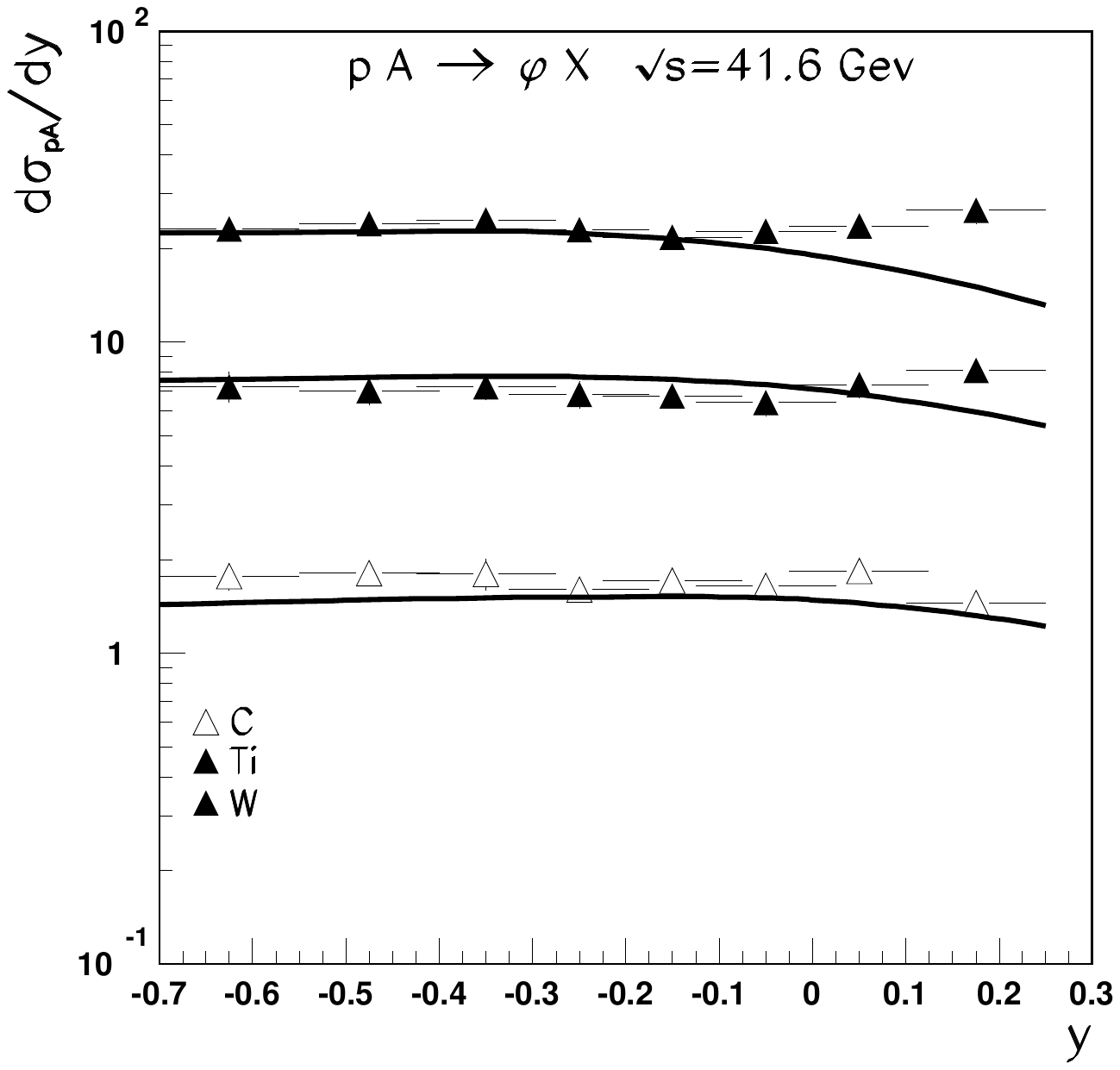}
\vskip -1.cm
\caption{\footnotesize
The experimental data on the $y$-spectra $d\sigma/dy$ of
$\varphi$-mesons produced in proton-nucleus collisions on different
($C$, $Ti$, and $W$) nucleus at $\sqrt{s}$ = 41 GeV~\cite{HERAb},
compared to the corresponding QGSM calculations.}
\end{figure}

\section{The $\varphi$-meson production in heavy ion collisions up to RHIC energies}

In the case of production of such a particles as pions and kaons, which give the main
contribution to the mean multiplicity at energies starting from $\sqrt{s}$ = 40$-$60~Gev,
new shadowing effects appear~\cite{CKTr} (see next section). On the contrary, in the case of
$\varphi$-mesons production these shadowing effects are absent,
even for the whole RHIC energy range, only appearing at LHC energies, as it will
be discussed in detail in the next section. 

Now, we will consider the $\varphi$-mesons production in heavy ion collisions
at energies from $\sqrt{s}$=17~GeV to $\sqrt{s}$=200~GeV.

The existing experimental data on midrapidity inclusive densities for produced $\varphi$-mesons
by the NA49 Collaboration ($\sqrt{s}$=17.3 Gev~\cite{NA49})
and those obtained at RHIC (STAR and PHENIX collaborations, $\sqrt{s}$=62.4~Gev~\cite{star2009},
$\sqrt{s}$=130~Gev~\cite{star130,star130adl}, and 
$\sqrt{s}$=200~Gev~\cite{star2009,starphi}) are presented in Table~2.

\begin{center}
\vskip 0.5cm
\vskip -10pt
\begin{tabular}{|c|c|c|c|c|} \hline
Reactions & Centrality & Energy & Experimental data & QGSM \\
%\cline{3-5}
% particles &CMS Collaboration & QGSM  & QGSM &  \\
   &  &  $\sqrt{s}$~GeV & $dn/dy$, $|y|\leq 0.5$  &  \\
\hline
%%%\hline

Pb + Pb & 0$-$5\%  & 17.3 & 2.35 $\pm$ 0.15, \cite{NA49}  & 2.764 \\ \hline

Au + Au & 0$-$20\% & 62.4 & 3.52 $\pm$ 0.08 $\pm$ 0.45, \cite{star2009}  & 3.36 \\ \hline

Au + Au & 0$-$11\% & 130. & 5.73 $\pm$ 0.37, $\pm$ 0.57, \cite{star130,star130adl}  &6.15 \\ \hline

Au + Au & 0$-$5\% & 200. & 7.95 $\pm$ 0.11 $\pm$ 0.73, \cite{star2009} & 7.57 \\ \hline

Au + Au & 0$-$5\% & 200. & 7.70 $\pm$ 0.30, \cite{starphi} & 7.57  \\ \hline

%Pb + Pb & 0 - 5\% & 2760. & 13.8 $\pm$ 0.5 $\pm$ 1.7 $\pm$ 0.1 & ALICE %\cite{alicepb} & 13.57  \\ \hline
%

%Pb + Pb & minimum bias & 2760. &  & ALICE 
%\cite{alicepb} & 3.98 \\ \hline

\hline

%%%\hline
\end{tabular}
\end{center}
Table~2: {\footnotesize The experimental data on $dn/dy$, $|y|\leq 0.5$, 
of $\varphi$-mesons production in different central nucleus-nucleus collisions 
at different energies, together with the corresponding QGSM results.} 

The experimental energy dependence of $dn/dy$ (y=0) is presented in Fig.~4,
together with the result of the QGSM calculation.
The rather strong energy dependence of this inclusive density comes from the fact that
the $\varphi$-meson has a significant mass, so the minimal values of $x_{\pm}$ in Eq.~(3)
noticeably decrease when the initial energy increases. This leads to the corresponding increase
of the integration region in Eq.~(4), and, consequently, to the increase of the inclusive density
in the midrapidity region.

%\newpage
\begin{figure}[htb]
\label{phi158ys}
%%%\centering
\vskip -9.cm
\hskip 3.0cm
\includegraphics[width=.9\hsize]{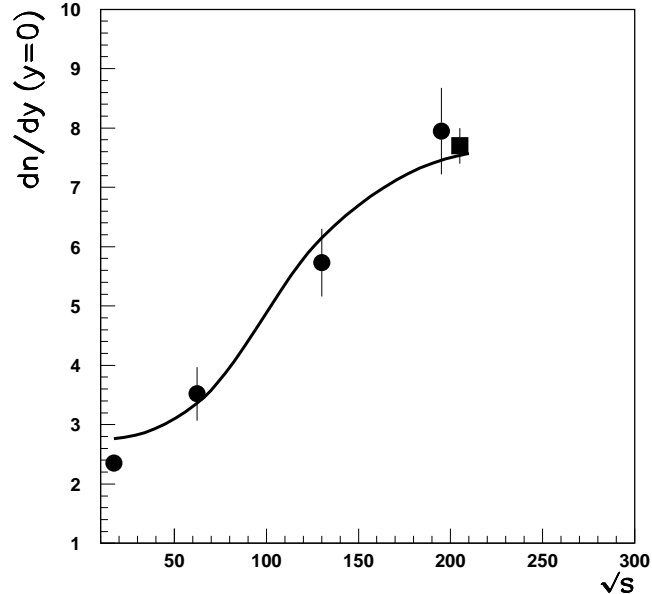}
\vskip -1.cm
\caption{\footnotesize
%Energy dependence of $\varphi$ meson inclusive density at midrapidity region 
%\cite{NA49,star2009,star130,starphi} together with the QGSM calculation.
The experimental energy dependence of inclusive density of $\varphi$-mesons produced
in the midrapidity region in both PbPb~\cite{NA49}
and AuAu~\cite{star2009,star130,starphi} collisions, together with
the corresponding QGSM calculation.}
\end{figure}

In Fig.~5 we compare the rapidity spectra $dn/dy$ of the produced $\varphi$-mesons in 
PbPb collision at 158 Gev/c~\cite{NA49} with the results of the QGSM calculations.
%%%Qualitatively,
In principle, the agreement at $y$=0 is rather good, but the theoretical curve seems
to fall down too fast with respect to the experimental data for $y>$ 0.   

%\newpage
\begin{figure}[htb]
\label{phi158y}
%%%%\centering
\vskip -9.cm
\hskip 3.0cm
\includegraphics[width=.9\hsize]{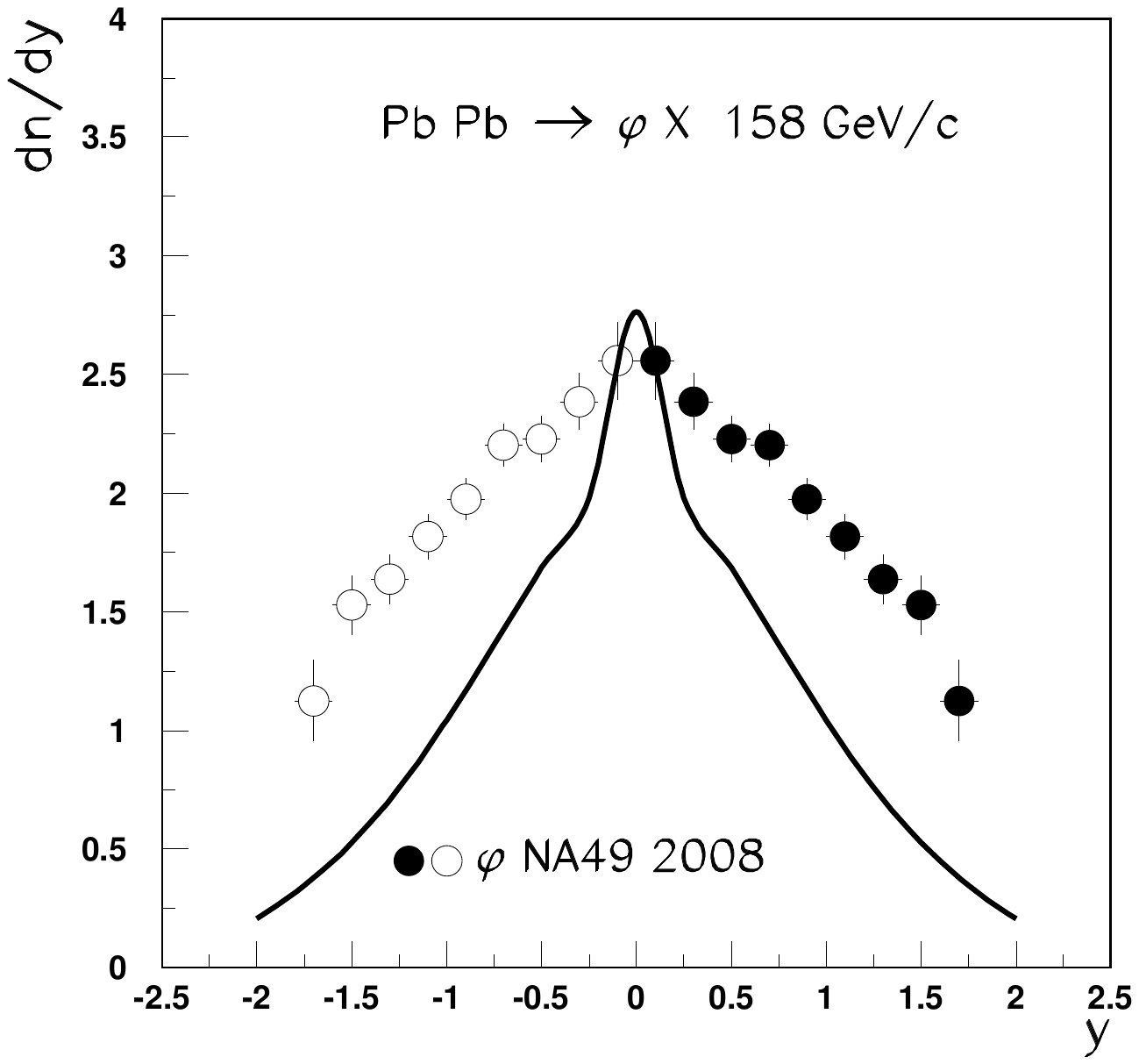}
\vskip -1.cm
\caption{\footnotesize
The experimental data on the rapidity spectra $dn/dy$ of
$\varphi$-mesons produced in PbPb collisions at 158GeV/c~\cite{NA49},
compared to the QGSM calculations.}
\end{figure}

\section{The $\varphi$-meson production at LHC energies}

In ref.~\cite{CKTr} it was explained that starting from RHIC energies
significant saturation effects for secondary production should be present
in both pPb and PbPb collisions, what has been since generally accepted, both
theoretically and experimentally~\cite{MPSd,Phob,Phen}.

At the same time, the spectra of secondaries produced in $pp$ collisions were
generally rather well described, without taking into account any saturation effects,
even up to the range of LHC energies.

These saturation effects can be explained by the inelastic screening
corrections connected to the multipomeron interactions~\cite{CKTr}, that
at low energies are negligibly small due to the suppression of the longitudional
part of the nuclear form factor. As this suppression of the longitudinal part of the
nuclear form factor decreases whith the growth of the initial energy,
the inelastic screening corrections become more and more significant
when the initial energy increases. 

The calculations of inclusive densities and multiplicities,
both in $pp$~\cite{CP1,CP2} and in heavy ion
collisions~\cite{CP2,CP3} (with accounting
for the inelastic nuclear screening),
can be performed in the percolation approach, and they result
in a good agreement with the experimental data for a wide energy region.

%%%The percolation model also provides a reasonable description
%%%of the transverse momentum distribution (at low and intermediate $p_T$),
%%%including the Cronin effect and the behavior of the baryon/meson
%%%ratio~\cite{Dias,Paj,Paj1}.
The percolation approach assumes that two or several Pomerons
overlap in the transverse space and they fuse in a single Pomeron.
Given a certain transverse radius, when the
number of Pomerons in the interaction region increases,
at least part of them may appear inside another Pomeron.
As a result, the internal partons (quarks and gluons) can split,
leading to the saturation of the final inclusive density. This effect
will persist with the energy growth, until all the Pomerons
will overlap~\cite{Dias,Paj,Braun}.

In order to account for the percolation effects in the QGSM,
it is technically more simple~\cite{MPSd} to consider in
the central region the maximal number of
Pomerons, $n_{max}$, emitted by one nucleon.
After they are cut, these Pomerons lead to the different
final states. Then, the contributions of all the diagrams
with $n \leq n_{max}$ are accounted for as at lower energies.
The unitarity constraint would also allow the emission of a
larger number of Pomerons $n > n_{max}$,
but due to fusion in the final state (on the quark-gluon
string stage) the cut of $n > n_{max}$ Pomerons would result in the same final
state as the cut of $n_{max}$ Pomerons.

With this prescription, the QGSM calculations become rather simple
and very similar to those in the percolation approach.
In this scenario, we obtain a reasonable agreement to
the experimental data on the inclusive spectra of secondaries at RHIC energy
with a value $n_{max} = 13$ (see ref.~\cite{MPSd}), and at LHC energies~\cite{MPSppb} with 
$n_{max} = 21$. The result that the number of strings for the secondary production
increases with the initial energy, even when the percolation effects are included, was
explained in ref.~\cite{JDDCP}.

It was shown in the previous section that in the case of $\varphi$-meson
production the inelastic shadowing effects are very weak at RHIC energy,
being not visible inside the experimental error bars. 
%It is agreement with the results of ref. \cite{bctpr} because the production
%of %$\varphi$- neson should be cross section of 
Thus, we can assume that the $\varphi$-meson is produced from the cut of a Pomeron with
relatively small transverse radius, so the overlapping and fusion of such a Pomeron
occur with small probabilities, and they become significant only at very high energies.
In such a picture, charm and beauty secondary particles would be produced from a Pomeron
with very small transverse radius (large transverse momenta of partons), so saturation
effects for charm and beauty secondaries should be many times smaller then for the $\varphi$-meson.
 
The experimental data for the inclusive densities of $\varphi$-mesons
produced in central PbPb collisions were measured by the ALICE Collaboration~\cite{alicepb}
at the energy $\sqrt{s}$ = 2.76 Tev, and they are presented in Table~3. In QGSM, we obtain
the value of $dn/dy$ ($|y|\leq 0.5$) = 13.8, in agreement with the experimental data, for the value
of $n_{max}$=37. The calculation with infinitely large $n_{max}$, i.e. without inelastic
screening, gives the value $dn/dy$ ($|y|\leq 0.5$) $\simeq$ 20.5, so the inelastic screening
effects for $\varphi$-meson production at $\sqrt{s}$ = 2.76 Tev estimated by the QGSM turns out to be
of $\simeq$ 1.5.

The value of the parameter $n_{max}$ was fixed as the result of the normalisation of 
QGSM calculations to the experimental data for $dn/dy$, $|y|\leq 0.5$.
The rapidity dependence of $dn/dy$ can be considered as the QGSM prediction.
This rapidity dependence is shown in Fig.~6 by solid curve. For comparison,
we also show our corresponding calculation without inelastic shadowing by a dashed curve.  

%\newpage
\begin{figure}[htb]
\label{phi276t}
%%%\centering
\vskip -9.cm
\hskip 3.0cm
\includegraphics[width=.9\hsize]{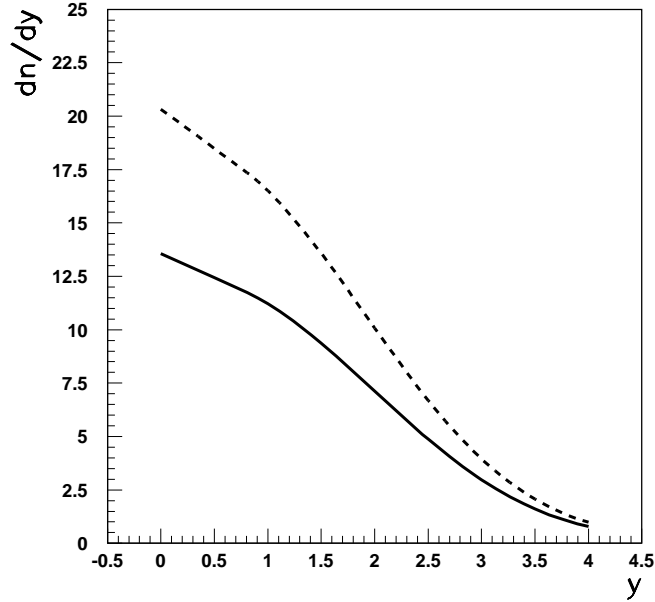}
\vskip -.8cm
\caption{\footnotesize
The QGSM prediction for the rapidity distribution of produced $\varphi$-mesons
in central PbPb collisions at $\sqrt{s})$ = 2.76 Tev, with (solid curve) and without 
(dashed curve) inelastic screening.}
\end{figure}

Once the value of $n_{max}$ is fixed, we can calculate the secondary $\varphi$-meson production
in proton-nucleus and nucleus-nucleus collisions, without any additional parameter with respect
to the corresponding calculations for pp collisions.
Thus, also in Table~3 we present the QGSM predictions for $\varphi$-meson production in
minimum bias PbPb collisions at $\sqrt{s}$ = 2.76~Tev, and in pPb collisions at $\sqrt{s}$ = 5~Tev.

\begin{center}
\vskip 0.5cm
\vskip -10pt
\begin{tabular}{|c|c|c|c|c|} \hline
Reactions &Centrality & Energy & Experimental data & QGSM \\
%\cline{3-5}
% particles &CMS Collaboration & QGSM  & QGSM &  \\
   &  &  $\sqrt{s}$~TeV & $dn/dy$, $|y|\leq 0.5$  &  \\
\hline
%\hline

Pb + Pb & 0$-$5\% & 2.76 & 13.8 $\pm$ 0.5 $\pm$ 1.7 $\pm$ 0.1 \cite{alicepb} & 13.57  \\ \hline

Pb + Pb & minimum bias & 2.76 & $-$ & 3.98 \\ \hline

p + Pb & minimum bias & 5.0 & $-$ & 0.16 \\ \hline

%%%\hline

%%%\hline
\end{tabular}
\end{center}
Table~3: {\footnotesize The experimental point on $dn/dy$, $|y|\leq 0.5$, 
on $\varphi$-mesons production in PbPb collisions at $\sqrt{s}$ = 2.76~Tev,
compared to the corresponding QGSM result, and together with the QGSM predictions
for minimum bias production of $\varphi$-mesons on a Pb target.} 
 
\section{Conclusion}

Up to the RHIC energies, the QGSM provides a reasonable description
of $\varphi$-meson production for the interactions of proton and nuclei
with nuclear targets, without any additional parameters with respect to the case of pp collisions.
The A-dependence of $\varphi$-meson production in proton-nucleus collisions presents
the usual behaviour $d\sigma_{pA}/dy (y=0) \propto A^1$, as it is shown in Fig.~2.
The dependence of the $\varphi$-mesons production in midrapidity region
on the initial energies is stronger than in the case of production of other hadrons.
Also the inelastic screening effects are weaker for $\varphi$-mesons production,
and they begin to be visible at higher energy $\sqrt{s}\geq $1~Tev.
Such a behaviour can be explained by the fact that the $\varphi$-mesons
are produced in cut Pomerons with relatively small transverse radii. From the point of view
of Reggeon Field Theory this means that the coupling of such a Pomerons to other Pomerons is weaker
than the coupling of a Pomeron with standard transverse radii. This effect should
be even stronger in the case of charm and beauty particle production, which are usually
described in the frame of perturbative QCD.

{\bf Acknowledgements}

We are grateful to M.G.~Ryskin for useful discussions. 

This work has been supported by Russian RSCF grant No. 14-22-00281,
by the State Committee of Science of the Republic of Armenia, Grant-15T-1C223,
by Ministerio de Ciencia e Innovaci\'on of Spain under project
FPA2014-58293-C2-1-P, and the Spanish Consolider-Ingenio 2010 Programme CPAN (CSD2007-00042),
and by Xunta de Galicia, Spain (2011/PC043).

%\newpage

\end{document}